\documentclass[11pt,letterpaper]{article}
\pdfoutput=1

\usepackage{graphicx,array}
\usepackage[dvipsnames]{xcolor}
\definecolor{darkblue}{rgb}{0,0.1,0.5}
\definecolor{darkgreen}{rgb}{0,0.5,0.2}
\definecolor{darkred}{RGB}{153,26,0}
\usepackage{ulem}
\usepackage{slashed}
\definecolor{seablue}{rgb}{0,0.2,0.6}
\usepackage{latexsym}
\usepackage{amssymb, amsmath}
\usepackage{slashed}
\usepackage{bm}        % for math
\usepackage[numbers,sort&compress]{natbib}
\usepackage{bm,relsize}
\usepackage{slashed}
\definecolor{viola}{RGB}{134,41,198}
\usepackage{mathrsfs}
\usepackage{hyperref} %Automatically links \label and \ref commands; Always load last
\hypersetup{
    colorlinks=true,       % false: boxed links; true: colored links
    linkcolor=darkblue,          % color of internal links
    citecolor=BrickRed,        % color of links to bibliography
    filecolor=magenta,      % color of file links
    urlcolor=MidnightBlue           % color of external links
}
\usepackage[all]{hypcap} %Link navagates to top of figure instead of caption (below fig)
\usepackage{subcaption}

\usepackage[font=small,labelfont=bf,labelsep=period]{caption}

\newcommand{\Tr}{\mathrm{Tr}}
\newcommand{\GeV}{\mathrm{GeV}}
\newcommand{\MeV}{\mathrm{MeV}}

\newcommand{\eV}{\mathrm{eV}}

\newcommand{\Mpl}{M_{\rm Pl}}

\newcommand{\SU}{\mathrm{SU}}
\newcommand{\be}{\begin{equation}}
\newcommand{\ee}{\end{equation}}
\newcommand{\PP}{\mathbb{P}}

 \date{\today}

\begin{document}

%%%%%%%%%%%%%%%%%%%%%%%%%%%%%%%%%%%%%%%%%%%%%%%%%%%%%%%%%%%%%%%%%%%%%%%%%%
\begin{flushright}

\end{flushright}
\vspace{.6cm}
\begin{center}
{\LARGE \bf 
Neutrinos, Dark Matter and Higgs Vacua \\
in Parity Solutions of the strong CP problem
}\\
\bigskip\vspace{1cm}
{
\large Michele Redi, Andrea Tesi}
\\[7mm]
 {\it \small
INFN Sezione di Firenze, Via G. Sansone 1, I-50019 Sesto Fiorentino, Italy\\
Department of Physics and Astronomy, University of Florence, Italy
 }
\end{center}

\vspace{.2cm}

\centerline{\bf Abstract} 
\begin{quote}
The strong CP problem can be solved if the laws of nature are invariant under a space-time parity exchanging the Standard Model with its mirror copy. We review and extend different realizations of this idea with the aim of discussing Dark Matter, neutrino physics, leptogenesis and collider physics within the same context. In the minimal realization of Ref.  \cite{Bonnefoy:2023afx} the mirror world contains a massless dark photon, which leads to a rather interesting cosmology. Mirror electrons reproduce the dark matter abundance for masses between 500-1000 GeV with traces of strongly interacting dark matter. This scenario also predicts deviations from cold dark matter, sizable $\Delta N_{\rm eff}$ and colored states in the TeV range that will be tested in a variety of upcoming experiments. We also explore scenarios where the mirror photon is massive and the mirror particles are charged under ordinary electromagnetism with very different phenomenology. We also show that, for the measured values of the SM parameters, the Higgs effective potential can give rise to a second minimum at large field value as required to break spontaneously the parity symmetry.
\end{quote}

\vfill
\noindent\line(1,0){188}
{\scriptsize{ \\ E-mail:\texttt{  \href{mailto:michele.redi@fi.infn.it}{michele.redi@fi.infn.it}, \href{andrea.tesi@fi.infn.it}{andrea.tesi@fi.infn.it}}}}

\newpage
\tableofcontents
\section{Introduction}

Discrete space-time symmetries, parity P, and time-reversal, T, are not fundamental symmetries of nature as we know it.
Parity is maximally violated by the chiral electro-weak interactions, and the presence of a complex phase in the Cabibbo-Kobayashi-Maskawa (CKM) matrix breaks T perturbatively.
In light of the CPT theorem, T violation is equivalent to CP violation and indeed all the observed CP violation observed in experiments is compatible with the one described by CKM matrix.
This leads to the so called strong CP problem because one can add the topological term to the QCD lagrangian
\be\label{eq:1}
\theta\frac{ \alpha_s}{16\pi} \epsilon^{\mu\nu\rho\sigma}G_{\mu\nu} G_{\rho\sigma}\,,
\ee
that breaks P and CP. Such term induces new CP violating effects in the Standard Model (SM), predicting in particular an electric dipole moment for the neutron $d_n\sim 10^{-15} \theta\, e$ cm.  
The experimental constraint $d_n< 10^{-26}\, e$cm  then implies that $\theta< 10^{-10}$, a value that appears inexplicable within the SM.

Essentially all the solutions of the strong CP problem rely on the existence of new symmetries.
In the axion solution (see \cite{DiLuzio:2020wdo} for a review) a U(1) global symmetry, known as Peccei-Quinn symmetry, is introduced that is anomalous 
under QCD but otherwise exact. Once the symmetry is spontaneously broken $\theta$ becomes a dynamical variable,
the axion, whose potential is minimized at the CP preserving point, thus solving dynamically the strong CP problem.
In the Nelson-Barr solution \cite{Nelson:1983zb,Barr:1984qx} CP is assumed to be an exact symmetry of nature.
If the symmetry is spontaneously broken in a favourable way an order one CKM phase can be obtained 
with suppressed $\theta-$term.

Similarly to Nelson-Barr models one might attempt to solve the strong CP problem imposing P as a fundamental symmetry that would immediately imply $\theta=0$.
At first sight this possibility is excluded because the SM is a chiral gauge theory that maximally violates P. Nevertheless it was realized long ago that P could be a spontaneously 
broken symmetry in extensions of the SM. In \cite{Babu:1989rb} a solution of the strong CP problem was proposed extending the electro-weak gauge sector. 
In \cite{Barr:1991qx} it was further shown that a full mirror electro-weak sector with common color interactions could give sufficient suppression to the $\theta-$ term.
The crucial observation was that the mirror sector should have mirror electro-weak interactions where SM fermions are also doubled but with opposite chirality as required
by parity, see also \cite{craig,Hisano:2023izx,Kuchimanchi:2023imj} for recent works.

Recently  an even simpler realization appeared in Ref. \cite{Bonnefoy:2023afx}. 
Here P is used in connection with a $Z_2$ symmetry that exchanges the SM with a mirror copy that is parity related,
\begin{equation}
\mathrm{\mathbb{P}[SM}]= \mathrm{\widetilde{SM}}^\dagger \,,\quad\quad \PP\equiv {\rm P} \times Z_2\,.
\end{equation}
Due to $\PP$ the mirror sector $\mathrm{\widetilde{SM}}$ has the same group structure of the SM and same matter content but representation are conjugated. 
In particular $\PP$ enforces the condition
\be
\theta_s\equiv\theta +\tilde\theta=0\,.
\ee
A simple solution of the strong CP problem then emerges where SU(3)$_c$ of the strong interactions is identified with the diagonal subgroup of SU(3)$\times \widetilde{\SU}(3)$ 
so that the coefficient of the topological term is $\theta_s$.  Crucially this solution is robust against  spontaneous breaking of parity.
The safest option that does not introduce new CP violating sources is to break $\PP$ spontaneously.
This can be done if the Higgs potential has a large field minimum so that in the mirror sector it has a large expectation value. 
Remarkably we find that with the addition of the new colored states a consistent minimum below the Planck scale 
emerges precisely around the observed values of the top quark and Higgs boson masses,  see \cite{Blinov:2016kte} for related work.

In this work we study cosmology and particle physics implications of the parity solution of strong CP. An important ingredient turns out to be neutrinos.
Neutrino masses can be generated through the see-saw mechanism that must be mirrored for consistency with $\PP$.
The symmetry allows to mix SM with mirror neutrinos that thus act as sterile neutrinos from the point of view of the SM.
The mirror neutrinos are often long lived and lead to entropy  injection in the SM thermal bath.
In the minimal scenario with massless mirror photon the mirror electron is stable an neutral under the SM and can be the Dark Matter (DM).\footnote{Similar ideas appeared in the literature in the context of Higgs parity theories \cite{Dunsky:2019api,Dunsky:2019upk,Hall:2019qwx,Dunsky:2023ucb}.}
If the abundance is determined by thermal freeze-out with the mirror photon we predict the DM mass in the range 500-1000 GeV. 
In this scenario we also predict a small fraction of color dark matter and contribution to $\Delta N_{\rm eff}$. Moreover DM is coupled to a long range force 
producing deviations from cold dark matter. We also explore scenarios where the mirror photon acquires mass. If the hypercharge is broken to the diagonal all the mirror states
have identical electric charge as the SM partners. This forbids a large reheating temperature and can produce deviation from the SM parametrized by
the $Y$  parameter \cite{LEP2}.

The paper is organized as follows. In section 2 we review the scenario of Ref. \cite{Bonnefoy:2023afx} and discuss generalizations
to different symmetry breaking patterns. A  study of the Higgs effective potential is presented showing that for the observed top quark mass 
consistent minima for the mirror world exist. We include right-handed neutrinos in section 3, discussing their lifetime and thermal leptogenesis.
We study the cosmology with a massless dark photon in section 4 showing that the scenario is compatible with large reheating temperature
and that DM can be reproduced by the mirror electron. Section 5 is devoted to the scenario with where hypercharge is broken to the diagonal.
We summarize our findings in 6.

\section{Solving the strong CP problem with $\PP$}

We start reviewing the main aspects of Ref.~\cite{Bonnefoy:2023afx}. The model has two exact copies of the SM related by the action of $\PP$. In particular, if $\psi$ is a left-handed Weyl fermion in a representation $r$ of SM, its $\PP$ conjugate would be a (left-handed fermion) $\tilde\psi$
\be
\PP[\psi]= \tilde\psi^\dag\,.
\ee
Since our definition of $\PP$ does not act on the representation of the gauge group, $\psi$ is in the representation $\bar r$ of $\mathrm{\widetilde{SM}}$.

If $\PP$ is a fundamental symmetry, the matter content of the model is completely fixed, as in table \ref{tab:UV}.
\begin{table}[t!]
\begin{center}
\begin{tabular}{ c | c | c | c }
\hline
SM & $SU(3)$ & $SU(2)$ & $U(1)$  \\ \hline 
$ Q$ & $ 3$ & 2 & $1/6$  \\ \hline
$ U$  & $ \bar 3$ & 2 & $-2/3$   \\ \hline
$ D$ &$ \bar 3$ &2 & $1/3$   \\ \hline
$ L$ & 1 & 1 & $-1/2$  \\ \hline
$ E$ & 1 & 1 & $1$   \\ \hline
\end{tabular}
~~
\begin{tabular}{ c | c | c | c }
\hline
$\mathrm{\widetilde{SM}}$ & $\widetilde{\SU}(3)$ & $\widetilde{\SU}(2)$ & $\tilde U(1)$  \\ \hline 
$\tilde Q$ & $\bar 3$ & 2 & $-1/6$  \\ \hline
$\tilde U$ & 3 & 1 & $2/3$   \\ \hline
$\tilde D$ & 3 &1 & $-1/3$   \\ \hline
$\tilde L$  & 1 & 2 & $1/2$  \\ \hline
$\tilde E$& 1 & 1 & $-1$   \\ \hline
\end{tabular}
\caption{\label{tab:UV} Quantum number of left-handed Weyl fermions of SM and mirror SM.}
\end{center}
\end{table}
In light of $\PP$ the lagrangian of the mirror sector is determined by the one of the SM as,
\be
\mathscr{L}_{\mathrm{\widetilde{SM}}}= \mathscr{L}_{\rm SM}(\PP[\psi,A], Y_{u,d,e}\to Y_{u,d,e}^*, \theta\to -\theta)\,.
\ee
From the behavior of the Yukawa matrices it is manifest that also the physical $\bar\theta$ angles of the two sectors are equal and opposite
\be
\bar{\theta}=\theta+\arg\det{Y_u Y_d}, \quad \bar{\tilde\theta}=\tilde\theta+\arg\det{Y_u^* Y_d^*}=-\bar\theta
\ee

In absence of right-handed neutrinos (that we discuss in detail in section \ref{sec:neutrinos}), the two sectors can communicate at the renormalizable level through the Higgs portal and kinetic mixing of the two hyper-charges, these two terms are invariant under $\PP$.\footnote{The presence of 
a kinetic mixing would have important implication for the scenario in \cite{Bonnefoy:2023afx} with a massless mirror photon.
Upon diagonalizing the kinetic terms the low  energy lagrangian can be cast in the form \cite{Fabbrichesi:2020wbt},
\begin{equation}
\mathscr{L}_{U(1)}=-\frac 1 4 F_{\mu\nu}^2-\frac 1 4 \tilde{F}_{\mu\nu}^2-e A_{\mu}\left(\frac {J_\mu}{\sqrt{1-\epsilon^2}}-\frac {\epsilon\,\tilde{J}_\mu}{\sqrt{1-\epsilon^2}}\right)-e \tilde{A} _\mu\tilde{J}^{\mu}\,.
\end{equation}
As a consequence the mirror matter acquires a charge $\epsilon$ under ordinary electromagnetism. Very strong bounds apply on this scenario 
especially if states of the dark sector are the DM. We will mostly focus on $\epsilon\approx 0$ in what 
follows.} 
As a boundary conditions at very high energies the couplings of the two sectors must be equal, but running to low energy can be different when $\PP$ is spontaneously broken, as we will discuss.  

A necessary ingredient to solve the strong CP is to identify our QCD as the diagonal combination of the two SU(3)'s of the visible and mirror sector.
This can be achieved introducing a scalar field $\Sigma$, transforming as a bi-fundamental $(3,\bar 3)$ or $(3,3)$ of the product group,
\be\label{eq:model}
\mathscr{L} = \mathscr{L}_{\rm SM} +  \mathscr{L}_{\mathrm{\widetilde{SM}}}- \frac{\epsilon}{2}B_{\mu\nu}\tilde B^{\mu\nu} -\lambda_{12}|H|^2|\tilde H|^2+  \Tr [(D_\mu \Sigma)^\dag D^\mu \Sigma] - V(\Sigma)\,. 
\ee
where the covariant derivative acts on $\Sigma$ as $D_\mu \Sigma= \partial_\mu \Sigma - i g_3 G_\mu \Sigma + i \tilde{g}_3 \Sigma \tilde G_\mu$. 

For a generic choice of the potential the minimum is realized for
\be\label{eq:diagonal}
\langle \Sigma_{ij}\rangle = \Sigma_0 \delta_{ij}\longrightarrow \frac {\SU(3) \times \widetilde{\SU}(3)}{\SU(3)_c}\,,
\ee
so that color groups are spontaneously broken to the diagonal that can be identified with the low energy QCD interactions.
At tree level the couplings satisfy the boundary condition
\begin{equation}
\frac{1}{g_s^2(M_\Sigma)}=\frac 1 {g_3^2(M_\Sigma)}+\frac 1 {\tilde{g}_3^2(M_\Sigma)}\,,
\label{eq:bcgauge}
\end{equation}
where $g_s(\mu)$ is the low energy QCD coupling and where we have loosely defined $M_\Sigma$ as the energy scale where $\SU(3) \times \widetilde{\SU}(3)$ is broken.
While other mechanisms of symmetry breaking exist, see \cite{Bonnefoy:2023afx} for examples,
many conclusions only depend on the symmetry breaking pattern. We will focus on the breaking 
with scalar bi-fundamental in the rest of the  paper.

The spontaneous breaking of SU(3) has two remarkable consequences. 
First, it solves the strong CP problem in the following way. The combination $(g_3 G_\mu \pm \tilde g_3 \tilde G_\mu)/\sqrt{g_3^2+{\tilde g}_3^2}$
acquires mass $m_G=\sqrt{g_3^2+\tilde g_3^2}\Sigma_0$. Being an octet under SU(3)$_c$ the phenomenology reduces to the one of colorons \cite{coloron1,coloron2}. 
The orthogonal combination corresponding to $g_3 G_\mu =\mp \tilde g_3 \tilde G_\mu$ is massless and can be identified with the gluons at low energies.
Due to $\PP$ one finds
\be\label{eq:matching}
{\cal L}_{\theta_s}=\theta\left[\frac{ \alpha_3}{16\pi} \epsilon^{\mu\nu\rho\sigma}G_{\mu\nu} G_{\rho\sigma} -\frac{ \tilde{\alpha}_3}{16\pi} \epsilon^{\mu\nu\rho\sigma}\tilde G_{\mu\nu} \tilde G_{\rho\sigma}\right]_{g_3 G_\mu =\mp \tilde g_3 \tilde G_\mu}=0\,.
\ee
Crucially the vanishing of the low energy $\theta$ angle holds even for $g_3\ne {\tilde g}_3$ so it holds even after the spontaneous breaking of $\PP$.
Second, due to eq.~(\ref{eq:diagonal}), the mirror quarks are colored under ordinary QCD at low energy. This implies that $\PP$ must be broken because otherwise the new colored states would have identical mass as the SM quarks. Experimentally the lightest mirror quark (the mirror up) must be heavier that 1.3 TeV \cite{Bonnefoy:2023afx}, and given the observed value of the up mass, this requires $\tilde v/v\gtrsim 10^6$ so that $\tilde v \gtrsim 2\times 10^8$ GeV. Concretely this is obtained when the mirror Higgs $\tilde H$ gets a VEV $\tilde v$ larger than $ v=246$\rm GeV. 
All the mirror fermion and gauge boson masses will be then lifted as compared to the SM by a factor of $\tilde v/v$. The mirror masses are thus
\be
m_{\tilde f} = m_f \frac {\tilde v}{v},\quad m_{\tilde W,\tilde Z}=m_{W,Z} \frac{\tilde v}{v}\,.
\ee
The lightest states of the mirror world beside neutrinos are the dark electron and dark up quark with masses.
According to lattice estimates the up quark mass is different from zero and in the $\overline{MS}$ scheme is found $m_u^{\overline{\rm MS}}(2\, {\rm GeV})\approx  2.3\pm 0.1$ MeV \cite{Fodor:2016bgu}, four times the electron mass. Taking in account running in a world where the electro-weak VEV is $10^ 8$ GeV the ratio changes roughly by factor 2 so that,
\begin{equation}
m_{\tilde u}\approx 2 m_{\tilde e} = 400\, {\rm GeV} \left(\frac {\tilde v}{10^8\, \rm GeV}\right)\,.
\end{equation}

We can distinguish two different scenarios,
\begin{enumerate}
\item $\Sigma_0 > \tilde v$\\
At energies  $E\lesssim M_\Sigma$ there are two copies of colored fermions of the SM.
This scenario is closely related to the one of \cite{Barr:1991qx} where a single gauge group SU(3) was considered. 
At energies  $E> m_\Sigma$ $\PP$ demands $g_3={\tilde g}_3$.
\item $\Sigma_0 < \tilde v$\\
For  $\Sigma_0 \lesssim 10^{-5} \tilde v$ the lightest color states are colorons. At energies greater than $m_\Sigma$ the symmetry is restored
but the couplings of the two sectors run differently due to breaking fo $\PP$. For $10^{-5} \tilde v \lesssim \Sigma_0 \lesssim \tilde v  $ some fermions are lighter 
$m_{\Sigma}$. $\PP$ demands that $g_3={\tilde g}_3$ at energies $E> \tilde v$.
\end{enumerate}

\subsection{Spontaneous breaking of parity}

As we discussed $\PP$ must be broken so that ${\tilde v} \gg v$.
The breaking can either be spontaneous or soft compatibly with the solution of the strong CP problem. 
We find it remarkable that the SM has the necessary ingredients to generate spontaneous breaking of $\PP$ because the Higgs potential 
develops a new minimum at large field values where the electro-weak symmetry is broken at a high scale.
For the observed value of SM parameters however in the SM the second minimum is typically trans-Planckian so it is likely beyond the regime 
of validity of the SM effective field theory. In the present setup new colored states exist so this conclusion must be 
reconsidered. Indeed we will find consistent vacua below the Planck scale.  Therefore the only assumption required for the parity solution to work is that the SM sits in the standard electro-weak minimum  while the mirror world lives in the large field minimum. This also guarantees that the breaking does not 
introduce new potentially dangerous phases.

The second minimum is modified by two effects compared to the SM. First the new colored states modify the evolution of the couplings and the effective 
potential above ${\rm Min}[m_{\tilde u}\,, m_{\Sigma}]$.
Moreover the Higgs portal coupling $\lambda_{12}$ modifies the potential at tree level. We can always tune the electro-weak minimum to the observed value but
then the second minimum will be predicted in terms of the SM couplings and it will be different from the one computed in the SM.
Note that even if $\lambda_{12}$ is set to zero it will be generated by the running so that in general the potential is  $V(H, \tilde H)$.
$\lambda_{12}$ is however induced only at high loop order so that $\lambda_{12}=0$ is in practise consistent.

The spontaneous breaking of a discrete symmetry leads to existence of topologically stable domain walls that would be disastrous if ever produced \cite{Kobsarev:1974hf} (see also \cite{Asadi:2022cco} for a recent discussion). For $\lambda_{12}=0$ the situation in our scenario is different. 
The domain walls interpolate between $(v,\tilde v)$ and $(\tilde v\,,v)$ that are exactly degenerate under exact $\PP$ symmetry. 
As long as $V=V(|H|^2) + V(|\tilde H|^2)$ the domain walls factorize into two independent profiles for $H$ and $\tilde H$ that are individually unstable
so that no domain wall problem arises. The reason why this happens is that there exists a family of unstable solutions that interpolate between different minima.
The presence of the Higgs portal is however expected to change this conclusion so that a single stable domain wall survives.
The domain wall problem could be eliminated if the $\PP$ symmetry is softly broken. Alternatively the domain wall problem is 
solved if $\PP$ is broken during inflation, $H_I< \tilde v$. In this case any abundance of domain walls is inflated away.

\subsection{Higgs effective potential at 2 loops: a second minimum before $\Mpl$}
In this section we discuss under what circumstances the SM Higgs effective potential can develop a sub-Planckian minimum compatible with the $\PP$ solution to the strong CP problem (see also \cite{Blinov:2016kte} for a different realization). We work in the limit of vanishing quartic portal $\lambda_{12}$. In this approximation and at the perturbative order of interest, the appearance of a mimimum at large Higgs vev for the mirror world can be studied by inspecting the SM effective potential, modified by the presence of new matter charged under the SM gauge symmetry as well as new dynamics.

The RGE running of the SM parameters is modified in several ways. First, the $\PP$ symmetry enforces equality of couplings at and above $\tilde v$. Second, importantly, above $M_\Sigma$ the strong coupling $g_s$ of our color SU(3)$_c$ is matched to the fundamental couplings of the two SU(3) factors as per eq.~\eqref{eq:bcgauge}, and this may happen before or after $\tilde v$. For example if $M_\Sigma \gg \tilde v$ the model becomes the one of Ref.~\cite{Barr:1991qx}. While in the opposite regime above $M_\Sigma$ the running of $g_3$ is modified by the presence of new colored matter.

\begin{figure}[t]
\centering
 \includegraphics[width=0.55\linewidth]{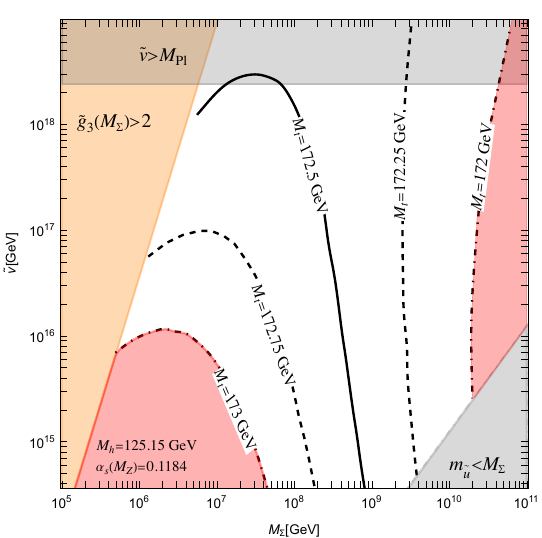}
\caption{\label{fig:solutions}\it   Parameter space of \cite{Bonnefoy:2023afx}  with spontaneous $\PP$ breaking. For the experimental value of the top quark mass, $m_t=172.5\pm0.5$ GeV \cite{CMS:2015lbj,ATLAS:2018fwq} a perturbative minimum below the Planck scale emerges for $10^6 \GeV< M_\Sigma < 10^{10} \GeV$. We assume  $\lambda_{12}=0$ and $M_\Sigma < m_{\tilde u}$.}
\end{figure}

\begin{figure}[t]
\centering\includegraphics[width=0.55\linewidth]{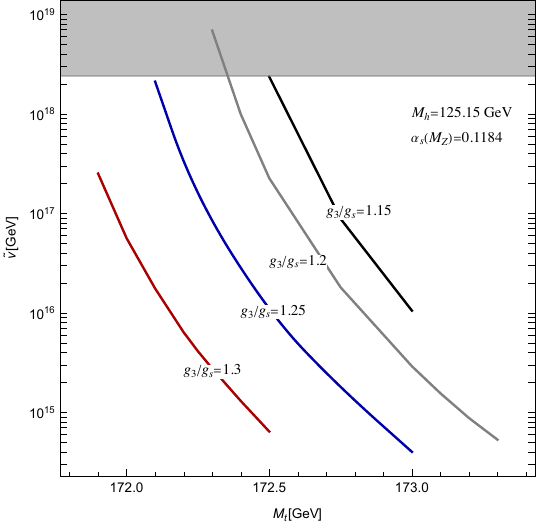}
\caption{\label{fig:vtilde}\it 
Correlation between the large field minimum of the Higgs and the top quark mass in the scenario  \cite{Bonnefoy:2023afx}.
Isolines are labelled by their corresponding value of $g_3(M_\sigma)/g_s(M_\Sigma)$. We assume  $\lambda_{12}=0$ and $M_\Sigma < m_{\tilde u}$.}
\end{figure}

We here explore the case where $M_\Sigma< m_{\tilde u}$ so that below the $\SU(3)\times \SU(3)$ breaking there is just the SM. The running of the SM parameter can be computed with great precision and we use \cite{Degrassi:2012ry} to run the $\overline{\mathrm{MS}}$ parameters up to the scale $\bar\mu=M_\Sigma$. Above this scale the running of $g_3$ is modified by the presence of the bifundamental complex scalar $\Sigma$. The modification of the $\beta$-function of $g_3$ at 2-loops reads\footnote{We neglect the 2-loop modification of top yukawa coupling and use the tree level condition (\ref{eq:bcgauge}).},
\be
\Delta \beta_{g_3}=\frac{g_3^3}{(4\pi)^2}\frac12+ \frac{g_3^5}{(4\pi)^4} 11
\ee 
The matching condition \eqref{eq:bcgauge} has the implication that the low-energy value of $g_3$ be limited to the range $g_s(M_\Sigma) \leq g_3(M_\Sigma) \leq \sqrt{2} g_s(M_\Sigma)$, since $\tilde g_3$ will be inevitably larger due to the faster running in the infra-red.

We then study the SM effective potential in this context, by fixing $M_t$, $M_h$ and $\alpha_s(M_Z)$ to their standard values, and inspecting for which values of $M_\Sigma$ a new sub-Planckian minimum for $h$ appears. The new physics acts as a stabilization for the Higgs quartic coupling: the matching condition will generate an effective larger $g_3$ coupling, that it is known to stabilize the SM potential, on top of the already smaller running due to the presence of $\Sigma$. Therefore, by fixing $M_t$ in the experimentally allowed range \cite{CMS:2015lbj,ATLAS:2018fwq} , $M_\Sigma$ should be large enough to allow for $g_3(M_\Sigma)$ to be such that a minimum right before $\Mpl$ appear. Intuitively, very large values of $g_3(M_\Sigma)$ will make the second mimimum disappear.

The solution $\tilde v(M_\Sigma)$ will be then very precisely determined upon a judicious choice of $g_3(M_\Sigma)$ and $\tilde g_3(M_\Sigma)$. This has to be determined iteratively satisfying both eq.\eqref{eq:bcgauge} and the $\PP$ symmetric boundary conditions at $\tilde v$, $g_3(\tilde v)=\tilde g_3(\tilde v)$. This is done by considering the running of $\tilde g_3$ from $\tilde v$ to $M_\Sigma$ including all the thresholds of mirror quarks (below $\tilde v$ the mirror sector is simply SU(3)$\times$U(1), we neglect the running of the mirror electromagnetism) up to three loops. We notice that the running of $\tilde g_3$ is much faster and can result quite easily in confinement of $\widetilde{\SU}(3)$, the running is faster since $M_\Sigma \ll m_{\tilde u}$. This poses a tight constraint to identify solutions with $\tilde g_3(M_\Sigma)$ in the perturbative realm.

The SM beta functions and effective potentials are taken from Refs \cite{Degrassi:2012ry,Buttazzo:2013uya}. New physics only enters through the boundary conditions at $M_\Sigma$ and the new contributions to the $\beta$-functions. Here we compute the effective potential for the Higgs as
\be
V_{\rm eff}(h)=\frac{\lambda_{\rm eff}(h)}{4} h^4\,,\quad\quad \lambda_{\rm eff}(\mu=h) = e^{4\Gamma(h)}[ \lambda(h) + \lambda^{(1)}(h) + \lambda^{(2)}(h) +\cdots]\,.
\ee
In order to compute this we use the available $\beta$-functions and the expression of the Higgs anomalous dimension, constructing a set of differential equations for $(g_i^2,y_i^2,\lambda,\Gamma)$. The effective potential then reads $V_{\rm eff}(h)=\frac14 \lambda_{\rm eff}(h)h^4$. 
The position of the second SM minimum is found by solving
\be
4\lambda_{\rm eff}(h) + \lambda'_{\rm eff}(h) h=0
\ee
By applying the algorithm discussed above we have found solutions with $\tilde g_3(M_\Sigma)$ in the perturbative regime, satisfying to great accuracy (sub-per-mille level) the boundary conditions and having a minimum $h<\Mpl$. Results are reported in figure \ref{fig:solutions} and 2. Notably consistent solutions are precisely found for the observed values of the top quark and Higgs mass.

In the scenario $M_\Sigma > \tilde v$  below the SU(3) breaking scale there are two copies of colored fermions. 
The running of $g_s$ is thus modified compared to the SM and is larger at high scales. This in turn reduces $y_t$ compared to the SM
and increases the instability scale until it disappears. In this case we have not found consistent minima with VEV smaller than 
the Planck scale and vanishing $\lambda_{12}$. Many other possibilities exist where only a subset of mirror fermions are lighter than $M_\Sigma$.
A more detailed analysis will appear elsewhere.

\subsection{Generalizations}
\label{sec:generalizations}

The breaking of the two fundamental SU(3)'s to the diagonal is a structural part of the scenario. One might wonder if more general patterns
of symmetry breaking can be considered. Focusing on two SM copies one can consider in general the pattern of symmetry breaking,
\be
\frac{[\SU(3)\times \SU(2) \times U(1)]\times [\widetilde{\SU}(3)\times \widetilde{\SU}(2) \times \widetilde{U(1)}]}{\SU(3)_c\times H}\bigg|_{\Sigma_0}\quad +  \quad [\widetilde{\SU}(2)\times \widetilde{U}(1) \to \widetilde{U}(1)_{\rm em}]_{\tilde v}\,
\ee
The model of Ref.~\cite{Bonnefoy:2023afx} corresponds to the case $H= \SU(2) \times U(1) \times \widetilde{\SU}(2) \times \widetilde{U(1)}$, where the electroweak interactions (visible and mirror) are unbroken by $\Sigma$.

Let us first consider the breaking $\SU(2)\times \widetilde{\SU}(2)$ down to the diagonal. In the regime $\tilde v \gg \Sigma_0$ mirror $\SU(2)$ is  broken
and one finds $v_{\rm SM}^2 = v^2 + \Sigma_0^2$. The small VEV of $\Sigma_0$ implies the existence of light 
vector bosons charged under $\SU(2)$ with mass in the 100 GeV range or below that is excluded. The opposite regime  $\tilde v \ll \Sigma_0$ is also not viable because
 $v_{\rm SM}^2 = v^2 + \tilde{v}^2$ so implying new light colored states. Therefore $\SU(2)\times \widetilde{\SU}(2)$ cannot be broken to the diagonal.
 This also forbids the possibility of unified breaking pattern of the form  $\SU(5)\times \widetilde{\SU}(5)\to \SU(5)_d$.

\bigskip

A different conclusion holds for hypercharge that can be broken to the diagonal combination without introducing new states with SM charges. We discuss this possibility below. 
Alternatively $\widetilde{U}(1)$ could be broken while preserving U(1), namely $H=\SU(2)\times U(1) \times \widetilde{\SU}(2)$. In order for this to be consistent with $\PP$ symmetry, one can for example add extra scalars charged under U(1)'s in both sectors or $\SU(2)$ triplets that spontaneously break only $\widetilde{U}$(1). Eventually, upon $\PP$ breaking, the only unbroken symmetries are the electroweak interactions of the SM. In this case the mirror states have the same quantum numbers of the model of Ref.~\cite{Bonnefoy:2023afx} but without massless mirror photon. 

\paragraph{Massive mirror photon}

\begin{table}[t!]
\begin{center}
\begin{tabular}{ c | c | c | c }
\hline
$q=0$ & $SU(3)_C$ & $SU(2)_L$ & $U(1)_Y$  \\ \hline 
$\tilde Q_u$ & $\bar 3$ & 1 & $0$  \\ \hline
$\tilde U$& 3 & 1 & $0$   \\ \hline
$\tilde Q_d$ & $\bar 3$ & 1 & $0$ \\ \hline
$\tilde D$& 3 &1 & $0$   \\ \hline
$\tilde L_e$ & 1 & 1 & $0$  \\ \hline
$\tilde E$& 1 & 1 & $0$   \\ \hline
$\tilde \nu$ & 1 & 1 & 0  \\ \hline
\end{tabular}
~~
\begin{tabular}{ c | c | c | c }
\hline
$q\neq 0$ & $SU(3)_C$ & $SU(2)_L$ & $U(1)_Y$  \\ \hline 
$\tilde Q_u$ & $\bar 3$ & 1 & $-2/3$  \\ \hline
$\tilde U$& 3 & 1 & $2/3$   \\ \hline
$\tilde Q_d$ & $\bar 3$ & 1 & $1/3$ \\ \hline
$\tilde D$& 3 &1 & $-1/3$   \\ \hline
$\tilde L_e$ & 1 & 1 & $-1$  \\ \hline
$\tilde E$& 1 & 1 & $1$   \\ \hline
$\tilde \nu$ & 1 & 1 & 0  \\ \hline
\end{tabular}
\
\caption{\label{tab:charges2} Mirror states quantum numbers below the symmetry breaking scale $m_{\Sigma}$ in the model with massless dark photon ($q=0$) 
and with hypercharge broken to the diagonal ($q\ne 0$).}
\end{center}
\label{tab:charges}
\end{table}

Breaking $U(1)\times \widetilde{U}(1) \to U(1)_Y$ corresponds to $H=\SU(2) \times U(1)_Y \times \widetilde{\SU}(2)$. This can be 
done effectively promoting $SU(3)\to U(3)$ and considering the collective breaking
\begin{equation}
\frac {U(3)\times \widetilde{U}(3)}{U(3)_d} \,.
\end{equation} 
We notice in fact that the scenario of \cite{Bonnefoy:2023afx} already contains the necessary ingredients and it is sufficient to 
to give $\Sigma$ a U(1) charge $(q\,,\pm q)$. The presence of the additional U(1) constrains the potential of $\Sigma$ of eq.~\eqref{eq:model} to be of the following form
\be
V(\Sigma)=-\mu_\Sigma^2 \mathrm{tr}[\Sigma\Sigma^\dag] +\lambda_\Sigma \mathrm{tr}[\Sigma\Sigma^\dag]^2  + \tilde\lambda_\Sigma  \mathrm{tr}[\Sigma\Sigma^\dag \Sigma\Sigma^\dag] ,
\ee
where we have absorbed the contribution from the Higgs vevs $v$ and $\tilde{v}$ in the effective mass term $\mu_\Sigma^2\geq 0$. If both quartics are positive the only minimum is the one leaving a U(3) invariant, $\Sigma\propto \mathbb{I}$ where $\Sigma_{ij}=\mu_\Sigma/\sqrt{3\lambda_\Sigma +\tilde\lambda_\Sigma}\delta_{ij}$.\footnote{Stability requires $\lambda_\Sigma +\tilde\lambda_\Sigma>0$. Possible unbroken subgroups compatible with stability are U(3) and U(2)$^2$ U(1). The former is the one discussed in the main text and it is realized for $\lambda_\Sigma\geq 0, \tilde\lambda_\Sigma > -3 \lambda_\Sigma$ or $ \lambda_\Sigma  > 0 ,\tilde\lambda_\Sigma>0$. The latter is the solution with $\Sigma_{ij} = \mu_\Sigma/\sqrt{\lambda_\Sigma +\tilde\lambda_\Sigma}\delta_{i3}\delta_{j3}$ for $\lambda_\Sigma>0 ,-\lambda_\Sigma< \tilde\lambda_\Sigma < 0$.
%\be
%U(3):\quad \, \sigma_1=\sigma_2=\sigma_3 =\sigma =\frac{\mu_\Sigma}{\sqrt{3\lambda_\Sigma +\tilde\lambda_\Sigma}}\,, \{\lambda_\Sigma\geq 0, \tilde\lambda_\Sigma > -3 \lambda_\Sigma\} or \{ \lambda_\Sigma  > 
%    0 ,\tilde\lambda_\Sigma>0\}
%\ee
%\be
%U(2)^2 U(1):\quad \, \sigma_1=\sigma_2=0,\quad \sigma_3 =\sigma =\frac{\mu_\Sigma}{\sqrt{\lambda_\Sigma +\tilde\lambda_\Sigma}}\,,\quad  \lambda_\Sigma>0 ,-\lambda_\Sigma< \tilde\lambda_\Sigma < 0
%\ee
}
In the unitary gauge the model of eq.~\eqref{eq:model} produces and effective mass term
\begin{equation}
\mathscr{L}_M= g_1^2q^2 \Sigma_0^2 (B-\tilde B)^2\,.
\end{equation}
Alternatively we could also give a Stueckelberg mass to the diagonal combination.

Due to the breaking to the diagonal the mirror states are charged under the SM hypercharge and thus also carry electric charge. 
In particular the mirror states have 
\begin{equation}
Q= Y = \tilde{T}_3 + \tilde{Y}=\tilde{Q}\,.
\end{equation}
Therefore all the mirror sector states have electric charge identical to the SM particles but with different masses
controlled by $\tilde v$. Note that instead the hyper-charge assignments of mirror matter (which is a good quantum number when $\langle H\rangle=0$) are different from the corresponding SM states. In this context the left-handed mirror neutrinos have vanishing hyper-charge (as well as, crucially, the VEV of the mirror Higgs field).

For $\Sigma_0 \to \infty$ the dark photon can be integrated out.
In this limit $B_1=B_2$, $G_1=G_2$ so that the gauge structure is,
\begin{equation}
SU(3) \times SU(2)_L\times SU(2)_R\times U(1)
\end{equation}
where $SU(2)_{L,R}$ act on SM and mirror fermions respectively. The same scenario was also considered in \cite{Barr:1991qx}.

\paragraph{More than one mirror}
The crucial ingredient of the strong CP solution is the convolution of space-time symmetry with mirror symmetry.
This can be generalized to 2N copies of the SM where,
\begin{equation}
\PP[\mathrm{SM}_i]= \overline{\mathrm{SM}_{i+1}} 
\end{equation}
Breaking color to the diagonal subgroup,
\begin{equation}
\frac{U(3)_i \times U(3)_{i+1}}{U(3)}
\end{equation}
gives rise at low energy QCD + a tower of states analogous to extra-dimensions. 
Indeed one could realize such scenario in extra-dimensions with replicas of the SM related by chirality.

\section{Neutrinos from the mirror world}\label{sec:neutrinos}

The generation of SM neutrino masses introduces new structure in the model that must be compatible with $\PP$ invariance. 
In this section we consider a type-I see-saw with right-handed neutrinos. 
To give masses to all neutrinos one needs to add 3 copies $(N_i\,,\tilde{N}_i)$ of singlet Weyl fermions  transforming under $\PP$
as $N_i \to \tilde N_i^\dag$. We can define the parity states,
\be
N^i_{\pm}\equiv N^i \pm \tilde{N}^i \to \pm (N^i_{\pm})^\dag\,.
\ee 
We consider the lagrangian invariant under $\PP$ (see also \cite{Chacko:2016hvu} for a particular choice)\footnote{We are here neglecting off-diagonal terms of the form $\Delta_{ij} N_i\tilde N_j$ with $\Delta$ an hermitian matrix.},
\begin{equation}\label{eq:autostati}
\mathscr{L}_\nu=\alpha_{ij} L_i H N^j_{+} + \beta_{ij} L_i H  N^j_{-}  + \alpha_{ij}^* \tilde L_i \tilde H  N^j_{+}  - \beta_{ij}^* \tilde L_i \tilde H  N^j_{-}   -  \frac{M_{+,i}}{2}N^i_{+} N^i_{+} -  \frac{M_{-,i}}{2} N^i_{-} N^i_{-} +h.c.\,,
\end{equation}
where $M_{+}$ and $M_-$ are real and diagonal, while $\alpha$ and $\beta$ are generic complex matrices in flavor space.   
In the limit $\alpha=\beta$ and $M_+=M_-$ two independent fermion number symmetries emerge for the visible and the mirror sector 
that guarantee that the lightest fermion of each sector is stable.

In a more symmetric notation, the above lagrangian can be recast as
\be\label{eq:N-symmetric}
Y_{ij}L_i H F_j + \widetilde Y_{ij} \tilde L_i \tilde H F_j  - \frac{M_i}{2} F_i F_i + h.c., \quad\quad F_i \equiv (N_+^i,N_-^i)\,.
\ee
The $Y,\widetilde Y$ yukawa matrices are $3\times 6$, and they are parametrized as $Y=(\alpha,\beta)$ and $\tilde Y=(\alpha^*,-\beta^*)$. In this notation $\PP$ enforces the following relation
\be
\widetilde Y = Y^* \mathcal{I}, \quad \quad \mathcal{I}=\mathrm{diag}[1,1,1,-1,-1,-1]\,.
\ee
This parametrization will be useful discussing leptogenesis.

We now discuss the possible structures of the neutrino mass spectrum. Since we consider the limit where $\mathrm{max}[\alpha,\beta]v\ll \mathrm{min}[M_\pm, \mathrm{max}[\alpha,\beta]\tilde]$ the SM neutrinos are always majorana. We can distinguish two cases:

\subsection{Mirror Majorana neutrinos}\label{sec:majorana}
We start with the limit where $M_{\pm}$ are the heaviest masses, so that $N_\pm$ can be integrated out. They act as see-saw for both the visible and the mirror neutrinos, and they yield the following lagrangian
\begin{eqnarray}\label{eq:full-majorana}
\mathscr{L}&=& i \bar L \slashed{D}L + i \bar{\tilde{L}} \slashed{D} \tilde L+\frac12 A_{ij} L^i H L^j H +\frac12 A_{ij}^* \tilde L^i \tilde H \tilde L^j \tilde H + B_{ij} L^i  H \tilde L^j \tilde H  +  h.c.,
\end{eqnarray}
where $A_{ij}$ is a symmetric matrix and $B_{ij}$ is an hermitian matrix in flavor space. In terms of the fundamental parameters they are given by
\be
A_{ij}=(\alpha \cdot\frac{1}{M_+} \cdot\alpha^T)_{ij} +(\beta \cdot\frac{1}{M_-} \cdot\beta^T)_{ij} \,,\quad \quad B_{ij}=(\alpha\cdot \frac{1}{M_+}\cdot \alpha^\dag)_{ij} -(\beta \cdot\frac{1}{M_-} \cdot\beta^\dag)_{ij}\,.
\ee
A unitary rotation $U$ and a diagonal rephasing $P$ make the first two terms diagonal and real (and therefore equal). This amounts to the  transformations $L\to U P L$ and $\tilde L\to U^* P^* \tilde L$, that take the lagrangian into the form
\be\label{eq:full-majorana}
\mathscr{L}=i \bar L \slashed{D}L + i \bar{\tilde{L}} \slashed{D} \tilde L+\frac{\tilde m_i}{\tilde v^2} \delta_{ij} L^i H L^j H + \frac{\tilde m_i}{\tilde v^2} \delta_{ij}\tilde L^i \tilde H \tilde L^j \tilde H + \frac{\Delta_{ij} \tilde m_j}{\tilde v^2}L^i  H \tilde L^j \tilde H  +  h.c.
\ee
where $\tilde m_i$ is real $2 \tilde m_i/\tilde v^2 \delta_{ij}= [P\cdot U^T\cdot A \cdot U \cdot P]_{ij}$ and $ \Delta_{ij}\tilde m_j/\tilde v^2=[P U^T B U^* P^*]_{ij}$. The presence of six right-handed neutrinos guarantees that all the Majorana neutrinos (visible and mirror) are massive. 

For vanishing $\Delta$ the mass spectrum of mirror neutrinos is determined by the SM neutrinos,
\begin{equation}
m_{\tilde \nu_i}= m_{\nu_i} \frac {\tilde v^2}{v^2}\,.
\label{eq:numass}
\end{equation}
Therefore the mirror neutrinos are determined by the same neutrino mass ordering of the SM. The lightest neutrino mass is bounded  by $m\ll \sqrt{\Delta m^2_{\rm sun}}\approx 0.008\, \mathrm{eV}$ in the normal ordering (NO),  by $m\ll \sqrt{\Delta m^2_{\rm atm.}}\approx 0.05\, \mathrm{eV}$ for the inverted ordering (IO) and $m\gg  \sqrt{\Delta m^2_{\rm atm.}}\approx 0.05\, \mathrm{eV}$ quasi-degenerate ordering (QD). One finds,
\be
\begin{split}
\tilde m_{\rm light.}^{\rm normal} = \left[1.9\big|_{\rm NO}, 12\big|_{\rm IO} \right] \GeV\, \left(\frac{m_{\tilde u}}{\mathrm{TeV}}\right)^2 \\
\tilde m_{\rm light.}^{\rm inverted} = \left[12\big|_{\rm NO}, 12\big|_{\rm IO} \right] \GeV\, \left(\frac{m_{\tilde u}}{\mathrm{TeV}}\right)^2
\end{split}
\label{eq:mirror-majorana}
\ee
Importantly the masses are quadratic in the VEV while the masses of other particles grow linearly.
As a consequence for $m_{\tilde \nu}>10$ TeV the mirror neutrinos are not the lightest states in the mirror world.

For $\Delta \ne 0$ visible and mirror neutrino mix.
After $\tilde H$ gets a vev, the lagrangian can be written as
\be
\mathscr{L}= i \bar L \slashed{D}L + i \bar{\tilde{\nu}} \slashed{\partial} \tilde \nu+ \frac{\tilde m_i}{\tilde v^2} \delta_{ij} L^i H L^j H + \frac{\tilde m_i}{2} \tilde \nu^i \tilde \nu^i  + \frac{\Delta_{ij} \tilde m_j}{\sqrt{2}\tilde v} L^i  H \tilde \nu^j  +  h.c. + \frac{1}{\tilde v^2}[\bar{\tilde\nu}\gamma^\mu\nu\bar{\psi}\gamma_\mu \psi]\,
\ee
and the mixing scales as
\be\label{eq:mixmajorana}
 \theta_{ij} \sim \Delta_{ij} \frac{ v}{\tilde v}=2\times 10^{-6}\, \Delta_{ij}\left(\frac{\mathrm{TeV}}{m_{\tilde u}}\right)\,.
\ee
Integrating out the mirror neutrinos we get the final Weinberg operator for the SM neutrinos
\be
\mathscr{L}= i \bar L \slashed{D}L + \frac{\tilde m_i}{\tilde v^2} \left[ \delta_{ij} + \frac{(\Delta \cdot \Delta^T)_{ij}}{4}\right] L^i H L^j H  +  h.c.
\ee
For $\Delta_{ij}\ll 1$, the neutrino masses are controlled by the high-scale see-saw, and their masses are strongly correlated with the mirror neutrino masses as in eq. (\ref{eq:numass}). Correlation is lost for $\Delta_{ij}\sim O(1)$ when the mass of observable neutrinos becomes dominated by the see-saw with the mirror neutrinos.

\paragraph{Mirror neutrinos decay}
Due to the spontaneous breaking of parity the mirror neutrinos $\tilde \nu$ are unstable towards decay to SM even if they are the lightest mirror fermions. 
The decay rate of $\tilde \nu$ is important for the cosmological history. Upon breaking of the mirror weak interactions, $\tilde L$ gets a majorana mass from the mirror Weinberg operator. In general, compatibly with $\PP$ a mixing exists between SM and mirror neutrinos, see eq.~\eqref{eq:full-majorana}. Such a term allows $\tilde \nu$ to decay even when electro-weak symmetry is unbroken. Clearly $\tilde \nu^i$ acts as three right-handed neutrinos for the SM neutrinos. The tree-level decay rate of the mirror neutrinos into $LH+L^*H^*$ is given by
\be
\Gamma(\tilde\nu_i \to \mathrm{all})\sim \frac{(\Delta^\dag \Delta)_{ii} }{16\pi}  \frac {\tilde m_i^3}{\tilde v^2}
\ee
Such a rate becomes faster than Hubble at a temperature 
\be
T_{\tilde\nu-dec}\lesssim 10^{3}\, \GeV \left(\frac {m_{\tilde u}}{\rm TeV}\right)^2 \left(\frac{m_i}{0.1 {\rm eV}}\right)^{3/2} |\Delta|^2\,.
\ee

Mirror neutrinos decay to SM final states from the $LH\tilde \nu$ term directly, or through off-shell $W/Z$ and Higgs. The decay width depends on the available decay channels. We here consider the case where the lightest mirror neutrino is below the $W$ mass, so that it decays to three fermion final states as well as lepton (charged or neutrino) plus meson final states below the QCD scale. The phenomenology here is totally analogous to a majorana 'heavy neutral lepton' (HNL), so that we can use results from the corresponding literature \cite{Chacko:2016hvu}. We notice that the decay to the lightest SM leptons and neutrino are of the form
\be
\Gamma_{\tilde \nu \to \mathrm{SM}}=C_F \frac{G_F^2 \theta^2}{192\pi^3}\tilde m^5 
\ee
Denoting with $m_{\nu, \rm light.}$ the mass of the lightest SM neutrino, the decay width is numerically given by
\be
\Gamma_{\tilde \nu \to \mathrm{SM}} \approx  2.8\times 10^4\,\mathrm{s}^{-1}\, C_F \left(\frac{m_{\nu, \rm light.}}{0.008\, \eV}\right) \left( \frac{m_{\tilde \nu}}{10 \GeV}\right)^4 \times \Delta^2\,,
\ee
where we assumed $\Delta_{ij}=\Delta \delta_{ij}$.
To determine whether $\tilde \nu$ is long-lived we can compare it to Hubble at $T\approx m_{\tilde \nu}$, their ratio reads
\be
\frac {\Gamma_{\tilde \nu \to \mathrm{SM}}}{H(m_{\tilde \nu})}\approx 1.4\times 10^{-4} \, C_F \left(\frac {90}{g_*(m_{\tilde \nu})}\right)^{\frac12}\left(\frac {m_{\nu, \rm light.}}{0.008 \, \eV}\right) \left(\frac {m_{\tilde \nu}}{10\,{\rm GeV}}\right)^2 \times \Delta^2\,.
\ee
From \cite{Chacko:2016hvu} we find $C_F \sim 16$ for $m_{\tilde \nu}\sim 10$ GeV.

\begin{figure}[t]
\centering\includegraphics[width=0.45\linewidth]{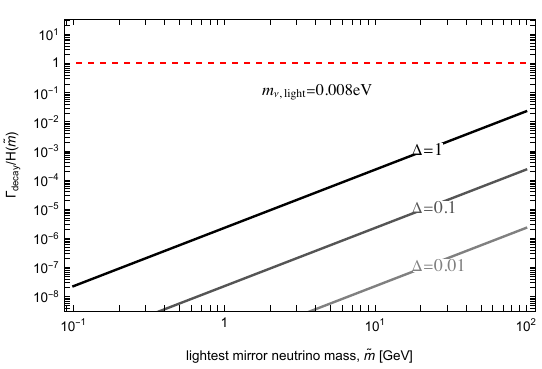}\quad \quad \centering\includegraphics[width=0.45\linewidth]{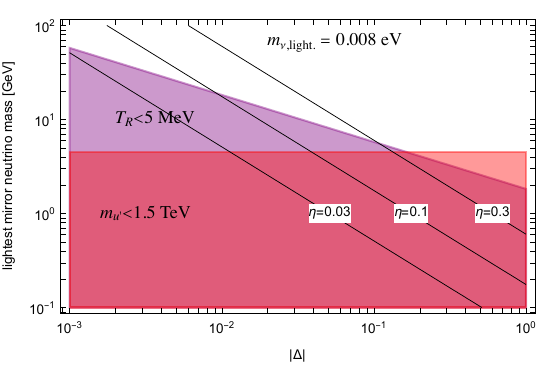}
\caption{\label{fig:nudecay}\it  On the left, lifetime of mirror neutrinos for the reference choice of SM neutrino $m_\nu= 0.008$ eV. On the right, isolines of dilution (\ref{eq:etadil})
are shown and bounds from BBN and mass mirror up quark.}
\label{fig:decay}
\end{figure}

For $\Gamma/H<1$ the neutrinos are long lived. Their energy density for $T< m_\nu$ redshifts non-relativistically until they decay to SM neutrinos when $\Gamma\sim H$.
If the decay is sufficiently slow the mirror neutrinos temporarily dominate the energy density of the universe injecting entropy in the SM thermal bath. 
This process dilutes all the abundance in the dark sector as well as the baryonic asymmetry. 
In order to avoid constraints from BBN the reheating temperature should be somewhat larger than 5 MeV. 
This translates into $\Gamma> 25/$s. The regions where neutrinos are long lived leading to entropy injection and 
are shown in blue in Fig. \ref{fig:nudecay} where the red region is excluded by BBN bounds.

Mirror neutrinos decouple after EWSB, at a temperature that can be computed by evaluating rates induced by Fermi operators with leptons (and light quarks). We call $T_{\tilde\nu,\rm dec}$ such a temperature, and mirror neutrinos decouples when relativistic. Their energy density at the time after they have become non-relativistic is 
\be
\rho_{\tilde \nu}(T)= \rho_{\rm rad}(T) \,\bigg[\frac{g_*(T_{\tilde m})}{g_*(T_{\tilde \nu, \rm dec})}\bigg]^{\frac13} \frac{\frac78 \times 2\times 3}{g_*(T_{\nu,\rm dec})} \frac{T_{\tilde m}}{T}
\ee
where $T_{\tilde m}\approx \tilde m$ is the temperature when they are non-relativistic. If  $\Gamma\ll H(T_{\tilde m})$, mirror neutrinos are long lived and they reheat the SM, thanks to an entropy injection. 

Most of the decay happens at a time around $\Gamma \approx H$, which might be already in a phase of matter domination driven by the mirror neutrino. At that time the corresponding SM temperature is
\be
T_\Gamma\approx 2.1 \bigg[\frac{g_*(T_{\tilde m})}{g_*(T_{\tilde \nu, \rm dec})}\bigg]^{\frac19} \left(\frac{g_*(T_\Gamma) \frac78 \times 2\times 3}{g_*(T_{\nu,\rm dec})}\right)^{\frac13}\left(\frac{\Gamma^2 \Mpl^2}{T_{\tilde m}}\right)^{\frac13}\,.
\ee
The maximal effect of entropy injection is when the neutrino decays while dominating the energy density of the universe. This gives the maximal amount of dilution, and correspond to the maximal reheating temperature
\be
T_R\approx  200\, \MeV   \Delta \sqrt{C_F} \sqrt{\frac {m_{\nu, \rm light.}}{0.008\, {\rm eV}}} \left(\frac {m_{\tilde \nu}}{\rm GeV}\right)^2\,.
\ee

The dilution is computed as the ratio of the SM entropy before and after the decay of the mirror neutrino. Using the simplified formulas in \cite{Garani:2021zrr} the dilution factor reads,
\begin{equation}
\eta_{\rm dil} \equiv \frac s {s_\Gamma}= {\rm Min}\left[ 1\,, \frac {1.3} {g_*^{1/4} Y_{\tilde \nu}} \sqrt {\frac {\Mpl \Gamma_{{\tilde \nu} \to SM}}{m_{\tilde \nu}^2}} \right]\approx {\rm Min}\left[ 1\,, 0.01\,   \Delta \sqrt{C_F} \sqrt{\frac {m_{\nu, \rm light.}}{10^{-2}\, {\rm eV}}}\left(\frac {m_{\tilde \nu}}{\rm GeV}\right)\right]
\label{eq:etadil}
\end{equation}
Note that the heaviest neutrinos whose mass should be larger than 10 GeV in light of the constraint on the mirror up quark
does not produce significant dilution. The dilution is thus dominated by the lightest neutrino.
BBN constraints require that the reheating temperature should be larger than 5 MeV or equivalently that $\Gamma \gtrsim 25\,{\rm s}^{-1}$. 
As a consequence only in the region around the physical values of neutrino masses the dilution is relevant. 
Dilution and BBN constrains are shown in \ref{fig:decay} right for the reference value $m_\mu=0.008$ eV.

\paragraph{Scattering of mirror neutrinos}
The same couplings that control the decay of mirror neutrino are also relevant for scattering at energies above their mass. Roughly the rate is given 
by the decay rate replacing $m_{\tilde \nu}$ with the temperature so that the rate goes as $T^5$. 
As a consequence the mirror neutrinos decouple when the temperature drops below the mass.

%%%%%%%%%%%%%%%%%%%%%%%%
%%%%%%%%%%%%%%%%%%%%%%%%
\subsection{Mirror Dirac neutrinos}\label{sec:dirac}
Now we investigate the opposite regime where the scale of the right-handed neutrino is below or comparable to the $\PP$ breaking scale, namely
where $\tilde v \gg M_{\pm}$ finding two different behaviors depending on the relative size of the two contributions. The discussion in this case will lead us to the vanilla see-saw for the SM neutrinos.

In this limit, the mirror neutrinos gets a Dirac mass term, $\tilde L^i \tilde H  (\alpha^*_{ij} N_+^j -\beta^*_{ij}N_-^j)$, that we can read out from eq.~\eqref{eq:autostati}. Below the scale of the Dirac mirror neutrinos, the effective lagrangian involving SM neutrinos and other states is found along the solution $\alpha^*_{ij} N_+^j=\beta^*_{ij}N_-^j$. 
Neglecting the flavor structure one finds,
\begin{equation}
\mathscr{L}_\nu =  \frac {2 {\rm Re}[\alpha^*\beta]}{\sqrt{|\alpha|^2+|\beta|^2}}L H N- \frac {(M_+ + M_-)(\beta^*)^2}{2(|\alpha|^2+|\beta|^2)}N^2 + h.c.
\end{equation}
Therefore, below the $\PP$ breaking scale, we are left with the standard scenario for the neutrinos where the right-handed neutrino is a combination of $N_+$ and $N_-$.
SM neutrinos are set by the see-saw scale, and we lose the correlation with the mirror majorana case.
In the case of 3 generations,  the Yukawa couplings are complex allowing for CP violation in the neutrino sector as required by leptogenesis.

\subsection{Leptogenesis}
The structure of right-handed neutrinos $N_\pm$ discussed above  allows us to discuss baryogenesis via leptogenesis, exploiting the asymmetric out-of-equilibrium decay of Majorana neutrinos as in standard leptogenesis (see \cite{Davidson:2008bu} and \cite{strumiaL} for reviews). In our context, we identify two different scenarios, that here we briefly sketch leaving further developments for future work.

\paragraph{$\PP$-broken leptogenesis}~\\
Below the scale $\tilde v$ a combination of $N_\pm$ acts as Majorana right-handed neutrino for the SM neutrinos (mirror neutrinos are instead Dirac). This falls into the category of standard leptogenesis. In this scenario the leptonic mirror sector plays no particular role since they are heavier than a Majorana combination of $N_\pm$. Therefore we can map this to the commonly discussed thermal leptogenesis. In particular, the interactions of the Majorana $N$ can be matched to the following (schematic) effective lagrangian
\begin{equation}
\alpha LHN + M NN + \frac{\alpha y_e}{\widetilde{M}_H^2} \tilde L N (\tilde L \tilde E)+ \frac{\alpha y_d}{\widetilde{M}_H^2} \tilde L N (\tilde Q \tilde D)+ \frac{\alpha y_u}{\widetilde{M}_H^2} \tilde L N (\tilde Q \tilde U)^* + (\alpha \to\beta)
\end{equation}
The branching fraction of $N$ decaying to mirror states of is suppressed by factors of order $y_{\rm SM}^2/4\pi^2 \times M_N^4/\widetilde{M}_H^4$, leading to $\Delta L +\Delta \tilde L\neq 0$ and $|\Delta L|\gg |\Delta \tilde L|$. We predict
\be
[B_0,L_0]=[\frac{-28}{79},\frac{51}{79}]\Delta L, \quad \quad [\tilde B_0,\tilde L_0]\approx 0\,, \quad\quad\mathrm{\PP-broken
}\ N-\mathrm{decay}\,.
\ee

\paragraph{$\PP$-symmetric leptogenesis}~\\
In this case the scale $M_N$ is the highest in the system, higher than $\tilde v$. This suggests that the decay of $N$ respects $\PP$. In this case both SM and mirror neutrinos are inevitably Majorana.

Since by symmetry, CP violation occurs equally in each sector, if the decay of $N$ happens when the mirror weak interactions are unbroken (therefore $\PP$ is not spontaneously broken) a net amount of lepton asymmetry in each sector will be produced. We refer to the lagrangian in eq.~\eqref{eq:autostati}
in the phase where $\PP$ is unbroken. The decays of $N_\pm\to LH,  L^* H^*, \tilde L \tilde H, \tilde{L}^* \tilde{H}^*$ are such that the produced number of $L$ is equal to the number of $\tilde{L}^*$. In light of this it is convenient to define the total generalized lepton number,
\begin{equation}\label{eq:totL}
N_{L_{\rm tot}}= N_L + N_{\tilde L}\,.
\end{equation}
Given that $\PP$ sends $N_L\to - N_{\tilde L}$ such number is odd under $\PP$, so that until $\PP$ is exact, the decay of $N_\pm$ will not generate any net $L_{\rm tot}$ number. Essentially, the second Sakharov condition is not satisfied for $L_{\rm tot}$. This can be checked explicitly by computing the asymmetry in the decay of $N$ in each sector,
\be
\varepsilon\equiv \frac{\Gamma(N\to LH)-\Gamma(N\to L^*H^*)}{\Gamma(N\to  L H)+\Gamma(N\to  L^* H^*)}\,,\quad \tilde\varepsilon\equiv \frac{\Gamma(N\to \tilde L\tilde H)-\Gamma(N\to \tilde L^*\tilde H^*)}{\Gamma(N\to \tilde L\tilde H)+\Gamma(N\to \tilde L^*\tilde H^*)}\,,
\ee
where $N$ is the heaviest neutrino. In order to compute the above asymmetries it is convenient to use the symmetric notation of eq.~\eqref{eq:N-symmetric}.
From an explicit computation we find (assuming $N=F_1$), 
\begin{eqnarray}
\varepsilon=\frac{1}{8\pi}\sum_{N_j\neq N} \frac{\mathrm{Im}[(Y^\dag Y)_{j1}^2]}{(Y^\dag Y)_{11}}f(\frac{M_j^2}{M_N^2})\,\quad \tilde\varepsilon=\frac{1}{8\pi}\sum_{N_j\neq N} \frac{\mathrm{Im}[(\widetilde Y^\dag \widetilde Y)_{j1}^2]}{(\widetilde Y^\dag \widetilde Y)_{11}}f(\frac{M_j^2}{M_N^2})\,,
\end{eqnarray}
where $f(x)=\sqrt{x}[\frac{x-2}{x-1}-(1+x)\log\frac{1+x}{x}]$ (see \cite{Davidson:2008bu} for a review). We notice that the loop function is symmetric under exchange of the two sectors, as it should. Moreover $\widetilde Y^\dag \widetilde Y= \mathcal{I} (Y^\dag Y)^* \mathcal{I}$. Notice also that the combination that enters the above expression is the square of $(Y^\dag Y)_{ij}$. We then have $(\widetilde Y^\dag \widetilde Y)_{ij}^2=(Y^\dag Y)_{ij}^{2*}$, and also noticing that $(Y^\dag Y)_{kk}=(\widetilde Y^\dag \widetilde Y)_{kk}$, yielding a relative sign between $\varepsilon$ and $\tilde\varepsilon$ upon taking the imaginary part we have explicitly $\varepsilon=-\tilde\varepsilon$, which supports the claim in eq.~\eqref{eq:totL}.

After $N$ decays and before mirror electroweak symmetry breaking an equal and opposite amount of lepton asymmetries is present in each sector, which can be transferred to baryon number through the respective sphaleron transitions, leading to equal and opposite baryon and lepton asymmetries in each sector. We notice that at this epoch both $B-L$ and $\tilde B-\tilde L$ are individually conserved. 

When the mirror Higgs develops a vev, mirror sphalerons shut down and $\tilde B$ becomes conserved. However $\tilde L$ violations become active again since the mirror neutrinos gets a majorana mass. Moreover, if $\Delta$ is sizable this can recouple $L$ and $\tilde L$. In turn, this can potentially lead to a washout of the SM baryon asymmetry produced thus far since both $B+L$ (SM sphalerons) and $L$ violating processes can now be active. This has the effect of (partially) washing out $B$ generated before. We leave this interesting scenario open to future investigations.

\section{Cosmology with a massless mirror photon}
\label{sec:masslessphoton}

The mirror sector contains colored states and a massless dark photon that  seem at odds with standard cosmology,
if the mirror sector is not empty. As we will show,  quite surprisingly the presence of a massless dark photon does not exclude the scenario and moreover the mirror electron 
elegantly produces the DM abundance from thermal freeze-out from mirror photons. The region of parameters selected by DM in particular is close to the experimental bounds.

The only link between mirror world and the SM is provided by colored states and neutrinos\footnote{We assume the kinetic mixing $\epsilon$ to vanish and the mixed quartic to be irrelevant.}.
For large reheating temperature the two sectors are in thermal equilibrium. Specifically if $T_R> \tilde v$
the whole dark sector including neutrinos thermalizes with the SM at a common temperature.
Neglecting for the moment the role of right-handed neutrinos, the first important event is the decoupling of mirror neutrinos, 
when electro-weak interactions of the type $\tilde \nu \tilde e \leftrightarrow \tilde \nu \tilde e$ go out of equilibrium. 
Mirror neutrinos, in absence of other dynamics, decouple relativistically at a temperature
\be\label{eq:mirror-nu-dec}
T_{\tilde \nu} \approx \left(\frac{\tilde v}{v}\right)^{\frac43} T_{\nu, \mathrm{SM}} \approx 5\times 10^4\mathrm{GeV} \left(\frac{m_{\tilde u}}{\rm TeV}\right)^{4/3}
\ee
If they are stable, they would overclose the universe given the masses (\ref{eq:mirror-majorana}).
The only viable option in this case would be that the reheating temperature is below the mass lightest colored state, $m_{\tilde u}$ or $m_\Sigma$,
so that the dark sector never reaches thermal equilibrium with the SM. 

A much more interesting interesting possibility is that the mirror neutrinos decay to the SM, $\Delta \ne 0$.
In this case the reheating temperature can be large and the two sectors are in thermal equilibrium initially.
Even if $T_R<T_{\tilde \nu}$ so that the neutrinos are not in equilibrium due to the mirror weak interactions,
the neutrinos portal will thermalize them. After the mirror neutrinos decay the dark sector contains massless mirror photon that contributes the dark radiation
and massive stable states, mirror electron and hadrons made of mirror up quark contributing to DM.

\subsection{Relativistic degrees of freedom}
SM and mirror sector maintain equilibrium until $T_*\approx {\rm Min}[m_{\tilde u}\,,m_\Sigma]/25$ through ordinary color interactions. Below $T_*$ the mirror sector contains mirror electrons and mirror photons, while at the same temperature the SM has its full thermal degrees of freedom. (After the mirror electrons decouple the mirror photon temperature gets reheated compared to the SM). 

Assuming that the two sectors decouple at temperature $T_*$ the evolution of the temperature is given in general by,
\be\label{eq:ratio-T}
\xi(T)\equiv \frac{\tilde{T}(T)}{T}= \left(\frac{\tilde{g}_*(T_*) }{g_*(T_*)}\right)^{\frac13} \left( \frac{g_*(T)}{ \tilde{g}_*(\tilde{T}(T))}\right)^{\frac13}
\ee
Assuming the mirror up quark to be the lightest colored state we estimate $T_* \sim m_{\tilde u}/25$.
At that temperature the only relativistic degrees of freedom in the mirror sector is the photon while mirror neutrinos typically remain into thermal
contact with the SM. We find,
\be
\xi(T\leq T_*)=0.34 \left(\frac{100}{g_*(T_*)}\right)^{\frac13} 
\ee
where is the effective number of degrees of freedom of the SM plus mirror neutrinos.
This allows us to compute $\Delta N_{\rm eff}$ from the presence of the light mirror photon, as
\be
\Delta N_{\rm eff}\big|_{\rm CMB}^4 =\frac{8}{7}(\frac{11}{4})^{\frac 43}\xi\big|_{\rm CMB}^4\approx 0.06 \left(\frac{100}{g_*(T_*)}\right)^{\frac43} 
\ee
While this contribution is currently consistent with experimental bounds it can be tested in future experiments . 

This result is modified if neutrinos are long lived. The entropy injection (\ref{eq:etadil}) modifies the previous results as,
\begin{equation}
\Delta N_{\rm eff} \approx 0.06  \eta_{\rm dil}^{4/3}
\end{equation}

\subsection{Thermal Dark Matter: mirror electrons (and mirror up quark)}

Mirror worlds contain automatically stable particles as a consequence of their accidental symmetries.
By construction mirror baryon number and electric charge are accidental symmetries of the mirror sector. 
It follows that the lightest state carrying baryon number and the mirror electron are stable\footnote{In the limit $\tilde v \to \infty$ mirror weak interactions decouple and individual fermion number
of leptons and quarks is conserved. In the relevant region of parameters the decay rate of heavier species, $\Gamma \sim M^5/{\tilde v}^4$ is fast enough to maintain thermal equilibrium with the plasma so the approximate stability does not play a role for the abundances.}. 
For the latter, since the U(1) symmetry is gauged and unbroken, stability is exact while for baryon number 
conservation holds up to dimension 6 operators as in the SM. 

As we will show the mirror electron is a good DM candidate being neutral under the SM. Its mass is predicted by the DM abundance and it turns out to be 
borderline with current constraints.  The lightest state with baryon number is the mirror up quark. This state is charged under ordinary QCD at low energy 
so naively it is not a viable DM candidate. The thermal abundance turns out to be suppressed compared to the mirror electron.
Moreover  as shown in \cite{DeLuca:2018mzn} heavy colored states mostly form deep Coulombian bound states that are SM singlets and are thus good DM components.
The residual small fraction of mirror up quark forms QCD-size exotic hadrons binding with ordinary matter.
These states feature large hadronic cross-sections and they are very constrained experimentally, their fraction should be smaller than $O(10^{-4})$.
Remarkably due to the double suppression of the abundance of hybrid QCD states the scenario appear to be viable. 
At the same time  deviations from CDM are predicted that might be observable in future experiments. 

\begin{figure}[t]
\centering
 \includegraphics[width=0.49\linewidth]{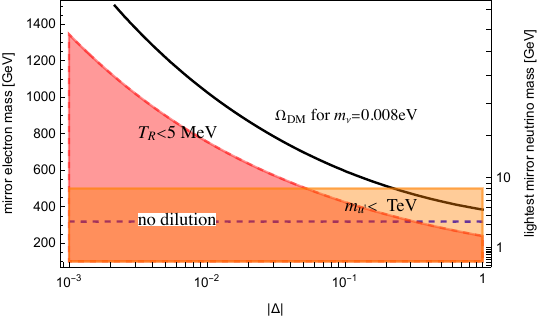}
 \includegraphics[width=0.49\linewidth]{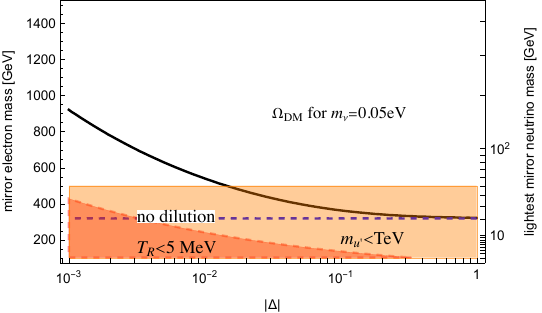}
\caption{\label{fig:abundance}\it  Mirror electron DM mass as  function of mixing $\Delta$ accounting for entropy injection (solid black line). On the left the abundance assuming that the relevant entropy injection is associated to the SM neutrino with mass $m_\nu= 0.008$ eV (normal hierarchy). On the right $m_\nu= 0.05$ eV (inverted hierarchy). Light red region is excluded by BBN constraints.}
\end{figure}

\paragraph{Mirror electrons as DM}~\\
The lightest charged states in the mirror world are electrons $\tilde e$ that are coupled to massless dark photons. 
The system at energies below $m_{\tilde u}$ is just dark QED,
\be
\mathscr{L}= \bar{\tilde e}i \slashed{D} \tilde e - m_{\tilde e} \bar{\tilde e}\tilde e -\frac 1 4 \tilde{F}_{\mu\nu}^2
\ee
Mirror electrons and up quark follow an equilibrium distribution due to annihilation into mirror photons.
Allowing for a different temperature between visible and mirror sector at decoupling one finds \cite{Garani:2021zrr},
\begin{equation}
\frac{\Omega h^2}{0.12} = \xi \frac {106.75}{g_*(T_*)} \sqrt{\frac{g_*(T_*)+\tilde{g}_*(T_*) \xi^4}{106.75}}\frac{1}{\langle \sigma_{\rm eff} v \rangle (23 {\rm TeV})^2}
\label{eq:abxi}
\end{equation}
where $g_*$ is the effective number of degrees of freedom of visible and mirror sector at freeze-out and $\xi$ the ratio of temperatures.

The tree level annihilation cross-sections of electrons is given by,
\begin{equation}
\langle \sigma_{e\bar e} v \rangle= \frac {\pi \alpha_{\rm EM}^2}{m_e^2}  
\end{equation}
This should be corrected by two related effects: and Sommerfeld enhancement and bound state formation that adds an annihilation channel 
 \cite{vonHarling:2014kha}. For $\alpha_{\rm EM}\sim 10^{-2}$ however the modification to the tree level cross-section is very small
and be neglected. 

At freeze-out $\xi\approx 1$ the temperature is between 10 and 50 GeV so that the pre-factor in eq. (\ref{eq:abxi}) can be neglected in what follows.
Computing the abundance with the tree level cross-section with $\alpha_{\rm EM}=1/128$ one finds that mirror electrons
reproduce the DM abundance for $m_{\tilde e}\approx  300$ GeV.
Using the fact that $m_{\tilde u} \sim 2 m_{\tilde e}$ we would thus predict  $m_{\tilde u}\sim 600$ GeV that is excluded experimentally.
This estimate  however neglects the slow decay of mirror neutrinos. If this happens for $T \lesssim m_{\tilde e}/25$ (the temperature where mirror electrons decouple
from the thermal bath) it leads to entropy injection in the SM thermal bath that the depletes the DM abundance according to the factor (\ref{eq:etadil})\footnote{We have checked that 
in the relevant regions of parameters freeze-out continues to take place in the relativistic regime.}. This effect is generic in most regions of parameter space.
We show the DM abundance  accounting for the entropy dilution in \ref{fig:abundance}. The critical abundance is reproduce for,
\begin{equation}
M_{\rm DM} = [300, 1000]\, {\rm GeV}
\end{equation}
allowing to evade possible collider bounds on colored states. 
A precise determination of the DM depends on the details of the neutrino sector and requires the solution of the coupled Boltzmann equations that
we leave to future work. 

There are two potential constraints on DM charged under a long-range interaction: $i)$ the delay of kinetic decoupling (suppressing power on small scales); $ii)$ the modification to DM properties in clusters and galaxies. The first problem is not very relevant for heavy DM. The kinetic decoupling can be estimated to happen $T_{k}\approx \mathrm{MeV} (M/\mathrm{TeV})^{3/2}/\xi^2 \times (0.01/\alpha) $ \cite{Feng:2009mn}, safely before BBN for heavy masses, while the second constraint extends  to even larger DM masses \cite{Agrawal:2016quu}. The strongest constraints arise from the structure of galaxy halos as for the observed ellipticity of the NGC720 halo \cite{ellisse}. Demanding that the interaction are sufficiently weak not to deform the DM velocity distribution over the lifespan of a galaxy one can estimate the bound \cite{Agrawal:2016quu}
\be
\alpha \lesssim 0.2 \left(\frac{M_{\rm DM}}{\mathrm{TeV}}\right)^2
\ee
so that mirror electrons in the TeV range can be good DM candidates.

Another constraint on this  scenario arise from underground direct detection experiments. 
As discussed in \cite{Dunsky:2019api,Dunsky:2019upk} even if absent at tree level a kinetic mixing between hyper-charges arises
at loop level. For $M_{\Sigma}> m_{\tilde u}$ running effects at 4-loop order generate $\epsilon \sim 10^{-8}$ that is already in tension with Xenon 1T. 
This conclusion assumes that the local DM density is not modified by the supernova shock waves that tend to expel it from the galaxy \cite{Dunsky:2018mqs}, 
see however \cite{McDermott:2010pa}. The effect is suppressed if $M_\Sigma\sim m_{\tilde u}$ as in that case the running only arises at higher loop level. 
This would however require a coincidence of scales. In any case direct detection experiments appear very relevant for this scenario and might already exclude this scenario.

\begin{figure}[t]
\centering
 \includegraphics[width=0.55\linewidth]{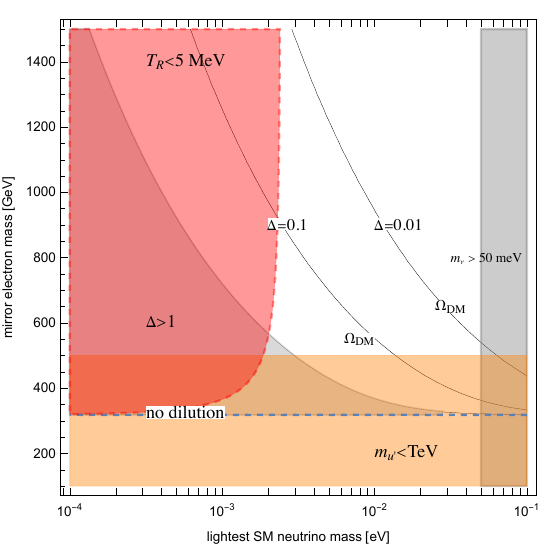}
\caption{\label{fig:summary}\it  Prediction for the SM lightest neutrino mass in connection with the DM relic abundance, for different values of $\Delta$. Light red region is excluded by BBN.}
\end{figure}

\paragraph{Colored DM fraction}~\\
The abundance of color triplets with no visible electric charge leads to a  component made of colored DM \cite{colorDM}.
Naively DM made of strongly interacting particles is not viable because for example it would form exotic nuclei where one of the ordinary quarks is  replaced
by the exotic quark. These nuclei have QCD like cross-sections and lead to very strong constraint on their abundance that can be at best a small fraction of DM. 
However for heavy DM, colored particles can form deep Coulombian bound states with binding energy $E_B \sim \alpha_3^2 M_Q$ much larger 
than the confinement energy. These bound states are small color singlets that behave as collision-less DM for practical purposes. The fraction of DM  that ends up into exotic  nuclei with QCD cross-sections and into Coulombian bound states is a dynamical question that requires to solve the coupled Boltzmann equations. As it turns out most of the DM actually
forms deep bound states that are energetically favoured leaving just  a small fraction of strongly interacting heavy quarks that hadronize with ordinary matter.

This general mechanism was explored in details in \cite{colorDM} for a scenario with a Dirac color octet. 
The QCD-like stable  bound states have in this case zero electric charge leading to weaker constraints. 
It was in particular shown that the abundance can be reproduced for a mass $M_{\tilde Q}=12.5$ TeV. 
In the current scenario the heavy quark is a color triplet with zero charge leading to exotic nuclei with fractional
electric charges. The experimental constraints are in this case stronger and it is unclear whether the scenario 
would be allowed if this made 100\% of DM. In our case however only a fraction of DM is in the form of colored particles leading 
to weaker constraints. Extremely strong bounds were however derived in  \cite{Dunsky:2018mqs} re-interpreting searches of magnetic monopoles
assuming that they are accelerated by supernova shock-waves and arrive to the detector with sufficient energy. 

The abundance from freeze-out can be simply estimated using the 
\begin{equation}
~~~~\langle \sigma_{u \bar{u}} v \rangle=\left[\frac 2 {43} S_{4/3}+\frac {41}{43}S_{-1/6} \right]\times  \frac {43}{54} \frac {\pi \alpha_3^2}{M_u^2} \,,~~~~~~~S_{\lambda}=\frac {2\pi \lambda \alpha_3/v_{\rm rel}}{1-e^{2\pi \lambda \alpha_3/v_{\rm rel}}}
\end{equation}
where $S$ is the Sommerfeld enhancement (computed in the massless limit) and we neglected the subleading decay to mirror photons.
Using $v_{\rm rel}\sim 0.3$ at decoupling and $\alpha_3(M_u)=0.1$ one finds an abundance of colored states $\Omega_{u\bar{u}}/\Omega_{e\bar{e}}\sim 0.1$.
We note that this estimate is conservative.  As discussed in \cite{colorDM} bound state formation enhances the effective annihilation 
cross-section reducing the abundance. Moreover after QCD confinement about half of the particles bind into unstable bound states 
leaving half of colored particles in states with net baryonic charge. Therefore the fraction of strongly interacting DM
is very suppressed.

In summary the scenario of \cite{Bonnefoy:2023afx} leads to sharp predictions for DM made of mirror electrons and up quarks with mass in the TeV range.
The scenario is currently close to experimental limit on DM charged under a long range force
predicting new colored states just above the collider constraints that give rise to a small component 
of strongly interacting DM. Future experiments can thus test this scenario.

\section{Phenomenology of massive mirror photon}

Let us now turn to the scenario where hypercharge as color is broken to the diagonal combination.
In this case a combination of U(1) gauge bosons has mass and the mirror states acquire SM quantum numbers as in the Table \ref{tab:charges2} right.
In particular the states have identical electric charge as the SM partners.\footnote{As discussed in section \ref{sec:generalizations} adding extra-dynamics such as electro-weak scalar triplets, the  photon could acquire mass in the mirror sector without breaking hypercharges  to the diagonal. In this case the mirror sector remains electrically neutral under the SM and the phenomenology can be similar to the one of \cite{Bonnefoy:2023afx} avoiding however possible constraints of DM charged under long interactions.}

To understand the phenomenology of the mirror photon in this context, it is useful to focus on the effective lagrangian of the neutral gauge bosons,
\begin{eqnarray}
\mathscr{L}_{\rm neutral}=&&-\frac 1 4 (W_{\mu\nu}^2 + \tilde{W}_{\mu\nu}^2 + B_{\mu\nu}^2 + \tilde{B}_{\mu\nu}^2+2\epsilon  B_{\mu\nu} \tilde{B}_{\mu\nu})\nonumber \\
&&+\frac {v^2}8 (g_2 W_{\mu}- g_1 B_\mu)^2 +\frac {\tilde{v}^2}8 (g_2 \tilde{W}_\mu- g_1 \tilde{B}_\mu)^2 + g_1^2q^2 \Sigma_0 ^2 (B_\mu -\tilde{B}_\mu)^2
\end{eqnarray}
where we neglect the difference of mirror couplings due to different running below $\tilde v$ and we also added a kinetic mixing for hypercharge as allowed by the symmetries. 

Given that for phenomenological reasons $\tilde{v}\gg v$, it is a good approximation to work in the limit $\tilde{v}\to \infty$. 
To leading order this corresponds to replace $\tilde W= g_1/g_2\times\tilde B$ in the above lagrangian so that,
\begin{equation}\label{eq:neutral-eff}
\mathscr{L}_{\rm neutral}^{\tilde{v}\gg v}= -\frac 1 4 (W_{\mu\nu}^2 + B_{\mu\nu}^2) -\frac 1 4 \left(1+\frac {g_1^2}{g_2^2}\right)\tilde{B}_{\mu\nu}^2- \frac{\epsilon}2  B_{\mu\nu} \tilde{B}_{\mu\nu} +\frac {v^2}8 (g_2 W_\mu- g_1 B_\mu)^2  + g_1^2 q^2 \Sigma_0^2 (B_\mu -\tilde{B}_\mu)^2
\end{equation}
The $3\times 3$ mass matrix can be exactly diagonalized since has a zero eigenvalue corresponding to the SM photon. 
However there are two distinct regimes depending on the relative size of $v$ and $q\Sigma_0$:

\paragraph{Light dark photon, $M_A\ll M_Z$}~\\
This corresponds to $q \Sigma_0\ll v$ so we can further expand around $v\to \infty$. 
Integrating out $W_\mu$ at tree level from \eqref{eq:neutral-eff}
one finds,
\begin{equation}
\mathscr{L}_{\rm eff}^{M_A\ll M_Z} = -\frac 1 4 \left(1+\frac {g_1^2}{g_2^2}\right)(B_{\mu\nu}^2+\tilde{B}_{\mu\nu}^2)-\frac {\epsilon}2 B_{\mu\nu}{\tilde B}^{\mu\nu} + q^2 \Sigma_0^2 (B_\mu -\tilde{B}_\mu)^2 -g_1 B_\mu J^\mu_{\rm EM}
- g_1 \tilde{B}_\mu \tilde{J}^\mu_{\rm EM} 
\end{equation}
where we have added the coupling to electro-magnetic currents. In terms of the mass eigenstates 
$\gamma= 1/\sqrt{2}(B+\tilde B)$ (massless photon) and $A= 1/\sqrt{2}(B-\tilde B)$ (the dark photon) the lagrangian takes the form
\begin{equation}
\mathscr{L}_{\rm eff}^{M_A\ll M_Z} = -\frac 1 4 \left(1+\frac {g_1^2}{g_2^2}+\epsilon \right)\gamma_{\mu\nu}^2-\frac 1 4 \left(1+\frac {g_1^2}{g_2^2}-\epsilon \right)A_{\mu\nu}^2 +2 g_1^2 q^2 \Sigma_0^2 A_\mu^2 -\frac{g_1}{\sqrt{2}} \gamma_\mu (J^\mu_{\rm EM}+\tilde J^\mu_{\rm EM}) -\frac{g_1}{\sqrt{2}} A_\mu (J^\mu_{\rm EM}-\tilde J^\mu_{\rm EM})
\end{equation}
For $E< m_A$ the SM and mirror electric currents couple identically to the massless dark photon: the mirror states have same electric charge
independently of $\epsilon$.  Note however that the coupling is reduced by $1/\sqrt{2}$ below and above $m_A$ so that QED is completely modified at low energies. 

In order to connect with the dark photon  literature it is useful a different basis.
We define $\gamma=B$ and $B-\tilde B=\sqrt{2}  A$ so that
\begin{eqnarray}
\mathscr{L}_{\rm eff}^{M_A\ll M_Z} &&= -\frac 1 2 \left(1+\frac {g_1^2}{g_2^2}+\epsilon\right)\gamma_{\mu\nu}^2 -\frac 1 2 \left(1+\frac {g_1^2}{g_2^2}\right)A_{\mu\nu}^2+ \frac 1 {\sqrt{2}} \left(1+\frac {g_1^2}{g_2^2}+\epsilon\right)\gamma_{\mu\nu}A_{\mu\nu}\nonumber \\
&& +2  q^2 \Sigma^2 A_\mu^2 -g_1 \gamma_\mu (J^\mu_{\rm EM}+\tilde J^\mu_{\rm EM}) +\sqrt{2} g_1 A_\mu \tilde J^\mu_{\rm EM}
\end{eqnarray}
In the language of dark photon this corresponds to an order 1 mixing.
Constraints on CMB $\mu-$distoritions require $M_A\lesssim 10^{-15}$ eV and even stronger bounds follow from Jupiter magnetic field, see \cite{Caputo:2021eaa} for a review.
For such small masses the phenomenology reduces to the  one of Ref. \cite{Bonnefoy:2023afx}.

\paragraph{Heavy dark photon, $M_A\gg M_Z$}~\\
In this regime from eq.~\eqref{eq:neutral-eff} the mass of the dark photon is $m_A^2\approx 4 q^2 \Sigma_0^2$, and it behaves as a perfect copy of the SM hyper-charge, since it couples to the hyper-charge current with coupling $g_1/\sqrt{2}$. Direct searches for this type on gauge bosons has been studied for example in \cite{Farina:2016rws,Alioli:2017nzr} with bounds in the multi-TeV region.

A heavy dark photon also leaves its imprints in terms of higher-dimensional operators in the SM effective theory. 
To study the precision physics effects on the SM it is convenient to integrate out $\tilde B$ and expand  in momenta,
\begin{equation}
\mathscr{L}_{\rm SMEFT}= -\frac 1 4 W_{\mu\nu}^2 -\frac 1 4 \left(2+\frac {g_1^2}{g_2^2}\right)B_{\mu\nu}^2 - \frac{\epsilon}2  B_{\mu\nu}^2 +\frac {v^2}8 (g_2 W_\mu- g_1 B_\mu)^2+ \frac {(g_1^2+g_2^2)^2}{4g_2^2 q^2 \Sigma_0^2}
(\partial_\rho B_{\mu\nu})^2
\end{equation}
From the kinetic term we extract the SM hypercharge gauge coupling,
\begin{equation}
\frac 1 {g_Y^2}= \frac {2(1+\epsilon)} {g_1^2}+ \frac 1 {g_2^2}
\end{equation}
The 4 derivatives term gives a contribution to the precision electro-weak parameter $Y$ parameter \cite{LEP12}
\begin{equation}
Y=\frac{2(1+\epsilon)^2 m_W^2}{m_A^2} \frac {(g_2^2+g_Y^2)^2}{(g_2^2-g_Y^2)}
\end{equation}
Current bound require $|Y|<0.2\times 10^{-3}$ \cite{ATLAS:2017fih} so that $m_A>10$ TeV
that is stronger than the one from direct searches. The bound is also expected to improve with future LHC data \cite{Farina:2016rws,Yfuture,WmassY}.

\subsection{Cosmological history}

Naively since the dark photon is massive  one might expect that this scenario would be less constrained.
In fact the opposite is true. Due to the spontaneous breaking of hypercharge to the diagonal all the mirror states have the same electric charge as the SM 
states. Since the mirror electron and up quark are stable in order to escape from bounds their abundance should be negligible. 
As we have shown if the mirror sector is initially in thermal equilibrium with the SM the abundance of electrons saturates the DM for masses below TeV.
The only possibility is thus that the mirror sector is never in thermal equilibrium with the SM. Given that the bridge between the two sectors is given by colored states we estimate,
\begin{equation}
T_R <\frac 1 {25} {\rm Min}[m_{\tilde u}\,, m_{\Sigma}]
\end{equation}
When this condition is satisfied the abundance of mirror is negligible. Since the reheating temperature is low the mechanisms thermal leptogenesis cannot be realized in this framework.
The only states that can be populated are thus neutrinos that are produced through the neutrino portal.

\paragraph{Mirror neutrinos as Dark Matter?}~\\
An interesting possibility for DM is offered by the mirror neutrinos that are the only neutral states when hypercharge is broken to the diagonal, see Table \ref{tab:charges2}. 
Mirror neutrinos with masses in the KeV range can give rise to sterile neutrino DM, see \cite{Abazajian:2017tcc,Boyarsky:2018tvu} for reviews.
Due to the mixing with the SM neutrinos they can be produced via freeze-in from the SM thermal bath.
The abundance is roughly reproduced for,
\be
\frac{\theta}{10^{-5}}\times \frac{m_{\tilde \nu}}{10\mathrm{keV}} \approx 1\,.
\ee
Note that if the mirror neutrinos are Majorana the mixing is given by (\ref{eq:mixmajorana})  so that it is difficult to obtain the critical abundance from freeze-in given the bound $m_{\tilde u}>1.5$ TeV. Other mechanisms could however be considered. Moreover in the regime where mirror neutrinos are Majorana the mirror neutrinos have masses given by eq. (\ref{eq:numass}).
This implies that the lightest neutrino must be lighter than $10^{-9}$ eV if the lightest mirror neutrino is in the KeV range. 

\section{Conclusions}

All known solutions of the strong CP problem rely on the existence of a new symmetry. 
In the case of the QCD axion the SM should be augmented with a $U(1)_{\rm PQ}$ global symmetry only broken by the QCD anomaly. 
The existence of an almost exact (apart from anomalies) global symmetry  raises the question of axion quality problem \cite{Kamionkowski:1992mf}, especially since in quantum gravity no global symmetries are known to exist. In the Nelson-Barr solution CP must be assumed to be an exact symmetry that is spontaneously broken and a non-generic structure of 
couplings of new fermions must also be enforced. 

From this point of view we find it fascinating that the strong CP problem might be solved by a space-time symmetry such as a generalized parity, $\PP$. 
This idea was originally advocated in the context of left-right models \cite{Babu:1989rb} and further generalized in \cite{Barr:1991qx} but has not received as much
attention as the QCD axion or even the Nelson-Barr mechanism. Here we have built upon Ref.~\cite{Bonnefoy:2023afx} where an even simpler scenario was considered. The SM is extended with a mirror copy related by spacetime parity. The breaking of ${\SU}(3)\times \widetilde{\SU}(3)\to \SU(3)_c$ then implies a boundary condition $\theta_s=\theta+\tilde{\theta}=0$ solving the strong CP problem.

In this work we  extended Ref. \cite{Bonnefoy:2023afx} in several ways.  First we have identified different patterns of symmetry breaking that allow to solve the strong CP 
problem but with different phenomenological implications. The existence of other minima in the SM Higgs effective potential suggests the possibility to break $\PP$ spontaneously.
We have shown that in the simplest scenario where the breaking of color is realized by a scalar bi-fundamental consistent high scale vacua exist precisely for the measured value of 
the top quark and Higgs masses. We have then included right-handed neutrinos necessary to give masses to visible and mirror neutrinos. 
This leads to an interesting structure of mirror neutrinos whose masses scale quadratically with the mirror Higgs VEV. The neutrino dynamics is particularly important 
for the cosmological history of the model as they are typically long lived. Moreover if the reheating temperature is large thermal leptogenesis can be realized in the SM without 
producing an asymmetry  in the mirror world. 

The phenomenology depends crucially on the fate of the mirror photon. If the mirror hypercharge remains unbroken as in \cite{Bonnefoy:2023afx}
all the mirror states have no electric charge under the SM. This leads to a very predictive scenario that is determined by the VEV of the mirror Higgs. 
Assuming a large reheating temperature the dark photon contributes to the effective number of relativistic degrees of freedom, predicting sizable $\Delta N_{\rm eff}$.
The mirror electrons reproduce the DM abundance for mass in the range 500-1000 GeV, that is marginally allowed by astrophysical and collider constraints and would produce signals in
direct detection experiments. A small fraction of strongly interacting DM is also predicted in the form of bound states of the mirror up quark that however leads to severe experimental constraints. 
The detailed predictions depend on the mirror neutrinos  whose decay leads entropy injection in the SM plasma. 
Overall the DM scenario appears somewhat in tension with experimental constraints motivating extensions of the minimal setup.

If the pattern of symmetry breaking also includes hypercharge so that $U(1) \times \widetilde{U}(1) \to U(1)_Y$ the mirror states have the same electric charge as the SM states. 
This implies that the mirror electron and up quarks that are stable have electric charge. As a consequence this scenario is only viable for a reheating temperature  lower than the mirror electron mass.  The only viable DM candidate is the lightest mirror neutrino that could realize sterile neutrino DM depending on the mass of the lightest neutrino. 
Direct and indirect collider constraints require that the dark photon mass is larger than 5-10 TeV.

All in all solving the strong CP problem through $\PP$ appears as an attractive possibility that should be considered seriously.
In this work we have just scratched the surface of the phenomenology of mirror $\PP$ world. 
A more detailed analysis will be surely needed to study the details of dark matter, leptogenesis and collider physics in this scenario. 
Many questions lie ahead that we are hoping to answer in the future. For example the spontaneous breaking  of parity gives a logic to the existence of 
a second minimum in the SM effective potential. It would be interesting to find a cosmological mechanism to populate the biverse.

\subsubsection*{Acknowledgements}
This work is supported by MIUR grants PRIN 2017FMJFMW and 2017L5W2PT. 
We would like to thank Daniele Barducci and Alessandro Strumia for discussions.

\pagestyle{plain}
\bibliographystyle{jhep}
\small
\bibliography{biblioCP}

\end{document}